\documentclass[10pt,a4paper]{article}
\usepackage{graphicx}
\usepackage{tabularx}
\usepackage{multicol}
\usepackage{tabto}
\usepackage{url}
\usepackage{enumitem}
\usepackage{textcomp}
\usepackage{eurosym}
\usepackage{amsmath}
\usepackage{MnSymbol}
\usepackage{wasysym}
\usepackage{xcolor}
\usepackage{booktabs}
\usepackage[labelfont=bf]{caption}
\usepackage[yyyymmdd]{datetime}

\usepackage{flowchart}
\usetikzlibrary{shapes,arrows.meta,chains,external}

\newcommand{\cz}[1]{\textit{\textbf{#1}}}
\newcommand{\eat}[1]{}

\newcommand{\ts}{
  \tikzset{>={Latex[width=2mm,length=2mm]}}
  \tikzstyle{line} = [draw, ->, >=latex, ultra thick]
  \tikzstyle{circ} = [
    circle,
    align=center,
    text width=3em,
    text centered,
    inner sep=0mm,
    outer sep=0mm,
    very thick,
    minimum width=0cm,
    minimum height=0cm
  ]
  \tikzstyle{box} = [
    draw,
    rectangle,
    ultra thick,
    align=center,
    text width=3.2cm,
    text centered,
    minimum height=2.5em
  ]
  \tikzstyle{nbox} = [
    draw,
    rectangle,
    thick,
    align=center,
    text width=3.6cm,
    text centered,
    minimum height=1.6em
  ]
  \tikzstyle{n} = [
    rectangle,
    align=left,
    text width=7.2cm,
    minimum width=1.5cm,
    minimum height=1cm
  ]
  \tikzstyle{asset} = [
    box,
    text width=2cm,
    align=center,
    color=white,
    fill=magenta!50!black
  ]
  \tikzstyle{noshape} = [minimum height=1em]
}

\begin{document}

\bibliographystyle{plain}
\thispagestyle{empty}

\begin{center}
\large{\bf Retail Central Bank Digital Currency:\\
Motivations, Opportunities, and Mistakes}\\
\end{center}
\vspace{0.5em}
\begin{center}
\begin{minipage}[t][][t]{0.3\linewidth}
\begin{center}
{Geoffrey Goodell}\\
\vspace{0.5em}
{\texttt{g.goodell@ucl.ac.uk}}
\end{center}
\end{minipage}
\begin{minipage}[t][][t]{0.3\linewidth}
\begin{center}
{Hazem Danny Al Nakib}\\
\vspace{0.5em}
{\texttt{h.nakib@cs.ucl.ac.uk}}
\end{center}
\end{minipage}
\begin{minipage}[t][][t]{0.3\linewidth}
\begin{center}
{Tomaso Aste}\\
\vspace{0.5em}
{\texttt{t.aste@cs.ucl.ac.uk}}
\end{center}
\end{minipage}
\end{center}

\begin{center}
{Future of Money Initiative}\\
\vspace{0.5em}
{Department of Computer Science}\\
\vspace{0.5em}
{University College London}
\end{center}
\begin{center}
{\textit{This Version: \today}}\\
\end{center}
\vspace{3em}

\begin{abstract}

Nations around the world are conducting research into the design of central
bank digital currency (CBDC), a new, digital form of money that would be issued
by central banks alongside cash and central bank reserves.  Retail CBDC would
be used by individuals and businesses as form of money suitable for routine
commerce.  An important motivating factor in the development of retail CBDC is
the decline of the popularity of central bank money for retail purchases and
the increasing use of digital money created by the private sector for such
purposes.  The debate about how retail CBDC would be designed and implemented
has led to many proposals, which have sparked considerable debate about
business models, regulatory frameworks, and the socio-technical role of money
in general.  Here, we present a critical analysis of the existing proposals.
We examine their motivations and themes, as well as their underlying
assumptions.  We also offer a reflection of the opportunity that retail CBDC
represents and suggest a way forward in furtherance of the public interest.

\end{abstract}

\section{Introduction}

The rise of the digital economy means that an ever-increasing share of retail
transactions are executed electronically, from e-commerce sales over the
Internet to in-person payments done via EMV terminals, contactless cards, and
mobile apps.  For the past thirty years, we have heard the argument that money
is memory, that credit is the consummate form of money, and that cash is only
really good for illegal activity and is, at best, an imperfect substitute for
credit.  And yet, digital currency is gaining salience at the same time that
consumers are increasingly relying upon bank money to engage with the economy
in general.  Cash allows consumers to conduct transactions without being
subject to the gaze of state or corporate surveillance, which can be used not
only to selectively block transactions but to inform lending, insurance, and
employment decisions.  Around the world, more businesses than ever are moving
their distribution online or refusing cash payments at the point of sale, while
at the same time, banks are choosing to close their branches and nudge their
depositors to make more of their payments online.

Information can always be copied, so it might be tempting to assume that
digital assets cannot be possessed and controlled the way physical assets can.
But this view is too simplistic: the way digital assets are used, they can
exhibit characteristics of possession and control.  Drawing from earlier work
by Goodell, Toliver, and Nakib~\cite{goodell2022}, and from published
standards, we suggest the following definitions:

\begin{itemize}

\item\cz{Possession:} property of a relationship between an actor and an asset
wherein that actor and no other actor can effect changes to the asset,
including its destruction, or transfer the asset to another actor

\item\cz{Control:} property of a relationship between an actor and an asset
wherein that actor and no other actor has the means to specify changes to the
asset, including the reassignment of this property itself to another actor,
that will be considered legitimate

\item\cz{Privacy:} ``the right of an entity (normally an individual or an
organization), acting on its own behalf, to determine the degree to which the
confidentiality of their private information is maintained''~\cite{iso24775}

\end{itemize}

We are now at a critical juncture in the history of money: Will people be able
to possess and control their own money and decide what to do with it, or will
all of their payment choices be subject to the oversight and permission of
intermediaries?  There is an opportunity to use digital technology to preserve
the privacy and property rights of everyday people and to frame the public
debate about how this will be done.

A tussle is emerging as a result of the changing nature of how the public uses
money and the multiplicity of vested interests and business motivations
involved in payments.  The design requirements for central bank digital
currency (CBDC) are currently under debate.  Around the world, various
proposals aiming to inform the design of digital currency have emerged.  In the
United States, the central bank conducted a consultation~\cite{fed2022}, and
Congress commissioned a report~\cite{crs2022}.  In the United Kingdom, the
central bank and treasury conducted a joint consultation~\cite{boe2023}
following other consultations by the central bank in previous
years~\cite{boe2020,boe2021}, and Parliament conducted a separate
consultation~\cite{lords2022}.  In the European Union, the central bank
conducted a consultation~\cite{ecb2021}, and the European Parliament published
a proposed regulation~\cite{ec2023}.  We shall critically review these
proposals in the next section.  Other relevant commentary and proposals have
been delivered by the central banks of other countries, including but not
limited to China~\cite{pboc2021}, Sweden~\cite{riksbank2020}, and New
Zealand~\cite{nz2021}, as well as by international organisations such as the
G7~\cite{g7}.  Through this debate, several areas of contention have emerged,
including but not limited to the following:

\begin{enumerate}

\item\cz{Privacy.} Consumer privacy is at risk in the digital economy.  Many
articles have identified the threat posed by a cashless society to the human
rights of individual persons, who must engage in the economy by making payments
as part of their everyday lives.  Indeed, it is trite to say that the purpose
of central banks is to ensure safe and equal access of public persons to the
economy.  Reports acknowledging the threats posed by electronic transactions
have been published around the world, by public-sector
organisations~\cite{allix2019,buttarelli2019,ca-opc2016,edpb2021,cnil2022}, by
private-sector businesses~\cite{mai2019,higgins2021}, and by civil rights
organisations and think tanks across the political
spectrum~\cite{naacp2019,gladstein2021}.  Central banks have also acknowledged
the dangers that cashless payments pose to
privacy~\cite{kahn2018,bindseil2023}, and the degradation of consumer privacy
is also understood to have significant implications for the broader
economy~\cite{garratt2021}.  Exceptional access mechanisms, which seek to
provide a means by which authorities can gain privileged access to data under
certain circumstances, have been consistently dismissed as dangerous and
untenable by the security community over the past quarter
century~\cite{abelson1997,abelson2015,benaloh2018}.

Data protection by trusted parties is recognised to be an imperfect substitute
for not collecting data linking consumers to their transactions in the first
instance~\cite{nissenbaum2017}.  Fortunately, alternative approaches are
available~\cite{rychwalska2021}, and the rise of the digital economy does not
imply that the relinquishment of privacy by individual consumers is inevitable.
In fact, the public utility of privacy in money has been
demonstrated~\cite{kahn2005}, and the response to a public consultation on CBDC
conducted by the European Central Bank revealed that privacy is the number one
concern of respondents~\cite{ecb2021}.  Some designs for CBDC demonstrate that
it is possible to offer true and verifiable privacy to
consumers~\cite{goodell2022,bis2022}.

\item\cz{Custody.} A salient feature of money is the option of individuals to
possess and control it directly as a bearer instrument, as they can with cash.
However, the custody model described in the relevant proposals in the
US~\cite{fed2022}, UK~\cite{boe2023}, and Eurozone~\cite{ec2023} assume that
consumers will not be able to possess or control CBDC.  Instead, CBDC would be
held in special ``wallets'', enabling the use of services offered by providers
who would be \textit{de facto} custodians of the money.  The implicit
definition of the term ``wallet'' as a \textit{service} contrasts with the
internationally accepted definition of a \textit{wallet} as an
``\textit{application or mechanism} to generate, manage, store or use'' digital
assets~\cite{iso22739} (emphasis added).

Direct custody of money underpins the free exercise of choice in market
economies, and if consumers cannot be in direct possession and control of their
money, then it is not really theirs.  Instead, they would be forced to contend
with the possibility that their use of money might be restricted to certain
purposes at certain times, as it was for recipients of the controversial
cashless welfare card in Australia~\cite{dss-au2022}, or curtailed completely,
as it was for Canadian lorry drivers in early 2022~\cite{woolf2022}, who were
subsequently exposed to public blacklisting~\cite{fraser2022}.

Requiring that intermediaries stand between consumers and their own money is
tantamount to requiring that consumers have money within custodial accounts.
We question whether, from the perspective of a consumer, ``digital currency''
residing within an account of this type is significantly different from bank
deposits residing within ordinary bank accounts.  We argue that digital
currency must be designed so that, like cash, it can be possessed independently
of an accounting relationship.

\item\cz{Role of identity.}  Separately from the question of custody, which is
about whether individuals can possess and control their own assets, is the
question of the form that the assets take.  Digital currency assets are
fungible, so in principle they can take the form of either \textit{balances}
(``account-based access'') or \textit{tokens} (``token-based
access'')~\cite{auer2020a}.  Both approaches are technically feasible: Some
cryptocurrencies, such as Bitcoin, conduct transactions using tokens, while
others, such as Ethereum, conduct transactions using balances.

However, the choice of balances versus tokens has implications for the role of
identity in accessing the assets.  In a digital currency system with balances,
a number representing the size of some collection assets must be reported
somewhere, and successive transactions must result in changes to that number.
The state of a balance, therefore, is determined by the net effect of a set of
transactions that are associated with that balance and with each other.  The
persistent linkage among transactions implies the existence of an
\textit{identifier} for the owner of the assets, as well as an implicit
relationship between this identifier and all of the transactions made by the
owner, as the owner of the assets must provide identification to access the
balance to conduct a transaction.  Such ownership can be economised but not
completely eliminated.  Conversely, with tokens, all that is needed to conduct
a transaction is knowledge of specific tokens or the cryptographic keys that
unlock them.  A token is not necessarily connected to other tokens, and
successive transactions made by the same person might not be linkable on the
ledger.  This is what happens with cash, which is, indeed, a kind of token.

The independence of a user's transactions is essential for both privacy and
control: Whether or not a particular gatekeeper has custody, the requirement to
use a balance implies that the balance can be locked as a mechanism for
preventing the owner from accessing the assets, or that links among successive
transactions can be forcibly discovered as a means of surveillance, as is the
case with the development of e-CNY system in China~\cite{goodell2021}.

\item\cz{Role of the central bank in processing transactions.}  Cash payments
are naturally decentralised; any two counterparties can consummate a
transaction without involving the issuer by exchanging a physical token.
Modern electronic payments are also decentralised; they are carried out by
networks of private-sector actors overseen by
regulators~\cite{bis2012,bis2012a}.

Trusting the central bank to operate (or delegate the operation of) a
transaction processing infrastructure is fundamentally different from trusting
the central bank to oversee a transaction processing infrastructure as a
regulator~\cite{goodell2021a}.  In particular, if a central actor were to
control the system that maintains the records of transactions, then there would
be no technical mechanism to stop this central actor from changing the
historical record.  The best we might expect is for changes to historical data
to be observable by periodic reconciliation activities among third parties with
read access to the records, followed by a legal challenge wherein the
authoritativeness of the records would be determined by indexing the intentions
of the maintainer of the records in some juridical context.  In the absence of
a clear procedure for determining which version of history is true, all parties
would be forced to accept the state of transactions decided by an authority,
irrespective of what actually transpired.  Even if we assume that the authority
is not corrupt, all participants in the system must depend upon the integrity
of its agents and the effectiveness of its security practices.  Recent history
teaches us that such dependence is a risky proposition~\cite{tucker2023}.

Finally, when the party that processes transactions also sets the rules, there
is nothing to prevent the rules from changing without warning.  Given how many
retail consumers would rely upon a future retail CBDC infrastructure, this risk
is too much to bear.  In distributed models for the operation of best-execution
networks, such as the National Market System in the United
States~\cite{nms-changes}, the regulator sets the rules but requires
participating private-sector actors to operate the system.  Changes must be
proposed, requested by the regulator, and implemented by all participating
actors before they can take effect.  This procedural approach protects the
system from hazards and is possible only because the system is decentralised in
practice.  Irrespective of consumer-facing service provision by intermediaries
or the potential use of privacy-enhancing technologies, many of the proposed
designs for CBDC assume that the central bank or some other centrally operated
entity would process individual retail transactions directly.

\end{enumerate}

These areas of contention feature prominently in the various proposals for CBDC
that have emerged in recent years, and they demonstrate that the future of
money is not simply a technical and economic issue, but fundamentally political
in nature.  The implementation of retail CBDC in economies with
well-established digital payments infrastructure would impact their payment
systems at different levels, from the technical underpinning of payments
infrastructure to the profitability of consumer banking, from geopolitical
aspects of money laundering regulation to the feasibility of delivering
economic stimulus packages.  Given the scale and scope of changes, it is
understandable that experts and non-experts alike might mistake the forest for
the trees.

Many of the concerns raised about the design of CBDC reflect an abundance of
caution, particularly relating to the use of CBDC to support criminal activity,
the potential to undermine the stability of the business models upon which
systemically important providers of financial services have come to rely, and
threats to the functions of central banks~\cite{bindseil2019}.  We shall engage
with each of these concerns carefully, with particular attention to the implied
trade-offs between different risks.  The rapid evolution of technology
sometimes means that the choice to do nothing is not be the conservative option
that it might otherwise seem to be.  Indeed, we argue that CBDC has been billed
as important for all the wrong reasons, many of which reflect a fundamental
misunderstanding of what CBDC could (or should) potentially achieve, how it
would be used, the role of central banking (or what it should be), and the
needs of consumers.

The remainder of this article is organised as follows.  In
Section~\ref{s:review}, we offer a review of proposals under consideration in
the United States, United Kingdom, and European Union.  In
Section~\ref{s:themes}, we identify and contextualise the main themes that
feature strongly in the proposals.  In Section~\ref{s:motivations}, we explore
the motivating factors behind why institutions and governments are spending
time and money on CBDC projects.  In Section~\ref{s:accounts}, we offer a
critical perspective on the state of public discourse on the nature of money,
focussing on underlying assumptions about accounts as the linchpin of the
relationship between retail payments and institutional power structures.  In
Section~\ref{s:opportunities}, we argue that as a new form of public money,
retail CBDC could be a democratising instrument that would promote prosperity,
enable competition, bolster innovation, and secure private property rights, but
only if it is implemented in a way that ensures that puts owners first.  In
Section~\ref{s:future}, we conclude, summarising the key challenges and
offering a path forward.

\section{A critical review of current proposals}
\label{s:review}

Next, we consider the design proposals for CBDC that have been developed by
various central banks, including the Federal Reserve System, representing the
United States; the Bank of England, representing the United Kingdom; and the
European Central Bank, representing the Eurozone in particular as well as the
European Union more broadly.  All of these central banks have championed
long-running initiatives to evaluate policy requirements for CBDC and to
develop a view on possible approaches to digital currency system design at a
technical level.  From their published proposals, we shall show that these
initiatives have embraced problematic implications for policy and have broadly
ignored important technical possibilities.  (We note that we have previously
reviewed the CBDC proposal of the People's Bank of China (PBOC), which shares
many of the same characteristics~\cite{goodell2021}.)  We ask whether these
initiatives have served the interests of incumbent actors in preference to the
broader public and nascent businesses, and if their constituents, that is,
everyday users participating in retail payments, have been abandoned.

\subsection{The United States of America: the Federal Reserve}

In January 2022, the Federal Reserve System produced a consultation paper,
``Money and Payments in the Age of Digital Transformation'', which describes a
set of design requirements for central bank digital currency~\cite{fed2022}.
Technology and system design are not policy-neutral, and although the authors
are careful to note that ``the paper is not intended to advance a specific
policy outcome''~\cite{fed2022}, the consultation paper nonetheless introduces
a set of assumptions, specifications, and technical characteristics that imply
a specific set of policy outcomes:

\begin{itemize}

\item\cz{Use of accounts.}  The consultation paper implicitly assumes that
individual users of CBDC would not hold it directly.  Instead, they would
``access CBDC'' via accounts with intermediaries.  Although never justified,
the assumption that CBDC must share its key functional features with bank
accounts, rather than cash, is intrinsic to the proposal.

\item\cz{Privacy.}  The consultation paper also suggests that privacy must be
limited to accommodate the ``transparency necessary to deter criminal
activity''~\cite{fed2022}, leaving open the question of who would decide
whether and how an appropriate balance would be achieved, presumably with the
assumption that the arbiters would have access to all of the data and decide on
the basis of what had been collected.  This treatment of privacy is consistent
with an article published in 2021 by the G7, which calls for ``balance''
between ``privacy and inclusion'' on one hand and ``reducing illicit finance''
on the other~\cite{g7}.

Unfortunately, because there is no way for individuals to verify how a set of
data is used once it has been collected, privacy ultimately depends upon
restricting the set of data that is collected in the first instance.
Therefore, depending upon the operational decisions of arbiters is problematic.
However, the authors did not justify their assumption that limiting privacy in
the interest of ``balance'' is necessary.  They failed to explore the nuances
of whether it might be possible to completely protect the privacy of some
parties but not others, or of whether it might be possible to collect some data
related to all transactions but not all.  Privacy and regulatory compliance are
and can be complementary, not contradictory, and specifically, it is possible
to design an architecture for digital payments that ensures tax and AML
compliance for recipients of money without collecting information that can be
used to link payers to information about how they spend their
money~\cite{goodell2022}.  However, the authors seem to wrongly assume that
solutions like this are impossible.

\item\cz{Intermediation and custody.}  The authors state that the initial
analysis by the Federal Reserve suggests that a CBDC system should be
``intermediated'', with the expectation that the private sector would offer
accounts or digital wallet services.  Despite the consideration for
non-custodial wallets offered by FinCEN~\cite{fincen2020}, the consultation
paper assumes that users would have no way to access CBDC except through
accounts maintained by intermediaries.  Thus, individuals would not be able to
possess and control their own money.  But there are other models for bearer
instruments that are not considered by the authors of the consultation paper,
who seem to assume that if a consumer is able to have possession and control of
a CBDC asset, then it must be possible:

\begin{enumerate}

\item to have peer-to-peer transactions without any involvement of third
parties;

\item to have a transaction without reporting the transaction to regulators;
and  

\item for both counterparties in a transaction to be anonymous.

\end{enumerate}

In fact, none of these assumptions are true.  The consultation paper ignores
the possibility that bearer instruments can be combined with transaction
infrastructure that directly enforces compliance rules.  It has been shown that
an architecture with self-validating tokens and oblivious transfers can be
built to allow consumers to directly possess the tokens that they use, while
also using blind signatures, as proposed by David
Chaum~\cite{chaum1982,chaum2021}, to make the payer anonymous.  The asset is
absolutely a bearer instrument in the sense that the payer holds a fungible
asset directly, does not provide identity information during transactions, and
is not subject to the rules of a custodian.  However, with a system of this
type, which is not considered in the consultation paper, it is possible to
combine irrevocable anonymity for payers with complete transparency for payees.

\item\cz{Role of identity.}  The authors assume that a requirement for
identification implies that transactions would be associated with a single,
unique identity.  Individuals would have identity-verified wallets, offered
subject to on-boarding procedures of the sort that might apply to bank accounts,
and that they would only be able to spend CBDC by identifying themselves to the
service provider at the time of the transaction.  The authors do not consider
that it might be possible to identify users at the time that they receive CBDC,
as they do when they receive cash from a bank, without requiring them to
identify themselves at the point of sale, as is the case with card payments.
In effect, the authors assume, without justification, that the transactions of
the future would be more similar to card payments, which require consumers to
identify themselves, than to cash, which does not have that requirement.

\item\cz{Holding limits.} The authors consider financial crises in which many
users simultaneously seek to switch from bank deposits to CBDC as a way to
reduce their exposure to risks associated with private-sector banks,
introducing the risk of bank runs and concomitant instability.  The authors
suggest that a central bank could ``address this risk by limiting the total
amount of CBDC an end user could hold''~\cite{fed2022}, although it is not
clear that holding limits imposed on wallets would be more effective than
limits on CBDC withdrawals in preventing rapid changes in the aggregate volume
of CBDC held by individual consumers.  Limits on withdrawals could be imposed
on individual consumers and businesses, perhaps with different limits for
different users, or they could be imposed across the board during crises,
perhaps as part of a rapid response mechanism.  It should also be noted that
limits to the total amount of CBDC in the system could be managed directly by
the central bank without monitoring individual transactions or individual
holdings at all, as is the case for cash.

\end{itemize}

We also note that at the time of the report, the Federal Reserve Bank of Boston
had already been engaged in Project Hamilton, a collaboration with the MIT
Media Lab following a line of technical research and design that had begun in
2016~\cite{bostonfed2022}.  Project Hamilton culminated in a software
implementation and analysis~\cite{lovejoy2022} that was subsequently presented
at NSDI 2023 in Boston~\cite{lovejoy2023}.  The salient features of this design
include a UTXO-based token management system wherein tokens are recorded and
spent using a distributed database with centralised control.  It is not private
by design, decision-making and the establishment of truth are entirely
centralised, and the database must capture all of the tokens and transactions.
However, the authors acknowledge that a token-based architecture is necessary
for privacy and that distributed ledger technology could be used to distribute
decision-making and control.  Although their proposed  architecture does not
use privacy-enhancing technology to isolate consumer identities from their
spending activities, the authors note that such features could be incorporated
into the system.  They even cite the longstanding work by Chaum, which
demonstrates that strong privacy outcomes for payers can be achieved even with
centralised architectures~\cite{chaum1982}, as well as his more recent work
with the Swiss National Bank on a design for central bank digital currency with
strong privacy for payers~\cite{chaum2021}.  However, the Project Hamilton team
did not incorporate these features into its design.

In December 2022, a group of nine members of the US House of Representatives
led by Representative Emmer of Minnesota penned an open letter to Susan
Collins, the President of the Federal Reserve Bank of Boston at the time,
questioning the engagement and role of private sector firms in Project Hamilton
and the implications of its vision for CBDC on the privacy of
individuals~\cite{emmer2022}.  Three weeks later, the Federal Reserve Bank of
Boston announced the conclusion of the project~\cite{lindsay2022}.  Meanwhile,
representatives on the other end of the political spectrum also raised concerns
about the potential for poorly-implemented digital currency to have deleterious
repercussions for society.  Representative Lynch of Massachusetts introduced a
bill promoting the development of a privacy-oriented electronic cash system
that would explicitly ensure that individuals would be able to spend money
anonymously~\cite{lynch2022}.  However, the bill presumed that the security
features would depend upon trusted hardware, which is problematic for many
reasons~\cite{goodell2022a}.

\subsection{The United Kingdom: the Bank of England}

In March 2023, the Bank of England and HM Treasury published a consultation
paper, ``The digital pound: a new form of money for households and
businesses?''~\cite{boe2023}, which is the latest document in a series that
includes ``Central Bank Digital Currency: opportunities, challenges and
design''~\cite{boe2020} in 2020 and ``New forms of digital
money''~\cite{boe2021} in 2021.  The consultation paper concerns the adoption
of central bank digital currency (CBDC) for retail use in the United Kingdom.
The consultation paper presents a detailed, prescriptive proposal for the
design of CBDC and includes strong positions on a variety of contentious points
concerning both policy and technology.  The positions demonstrate that the
authors had made some problematic assumptions, and the resulting system design
appears to favour the interests of certain stakeholders in preference to the
interests of the general public, particularly retail consumers.  We consider
several of the most salient issues below:

\begin{enumerate}

\item\cz{Motivation and process.}  Overall, although the consultation paper
references a variety of specific benefits of CBDC, it is broadly silent about
the underlying motivation for considering implementing CBDC in the UK.  Many of
the arguments presented throughout the consultation paper depend upon accepting
certain assumptions, for example that modern electronic payments are
well-suited to serve the interests of those who use them, that negate important
arguments for researching and developing CBDC, particularly the need to develop
an adequate analogue for cash in the digital economy.  The lack of clarity
about the motivation for developing CBDC undermines the strength of those
arguments and raises doubts about their validity.

\item\cz{Privacy.} The consultation paper states that transactions will not be
anonymous because identifying transacting parties is ``needed to prevent
financial crime''~\cite{boe2023}.  The privacy model for CBDC reflected in the
consultation paper relies upon third parties to collect and safeguard
transaction data, and it is assumed that methods would be available by
authorities to access such data.  Inevitably, although the Bank of England
itself might not be able to de-anonymise payers, other parties, including but
perhaps not limited to authorities, would be able to do so, and they would be
in a position to unilaterally judge whether data protection is appropriate or
not.

The consultation paper repeatedly asserts that a level of privacy similar to
that of card payments ought to be sufficient for CBDC, characterising the
digital pound as ``privacy protected like cards and bank accounts'', ``at least
as private as current forms of digital money, such as bank accounts'', ``on the
same basis as currently with other digital payments and bank accounts more
generally'', ``providing the same privacy as most of the money we use'', and so
on~\cite{boe2023}.  But payments via card networks or bank transfers are not
private, as anyone who has seen a bank statement can confirm.  Concluding that
this level of privacy is acceptable for public infrastructure simply because
most transactions have this level of privacy completely ignores the
requirements of the transactions that are in the minority, as well as the
legitimate concerns of those individuals who accept this level of privacy not
as the result of true consent but simply because there is no alternative.

The consultation paper also suggests business justifications for data
collection, arguing that intermediaries can be expected to ``use transaction
data to improve existing operations or to offer new customer-facing services''
and that there is ``public appetite for trading personal information for access
to products and services''~\cite{boe2023}.  Whether or not this is true, it is
not at all clear that the Bank of England should consider the support and
promotion of such business models to fall within its remit.

\item\cz{Custody.}  The consultation paper suggests that payment interface
providers (PIPs) would provide wallets, and although PIPs would ``never be in
possession of end users' digital pound funds'', all assets would be represented
directly on the core ledger, managed and discoverable by core ledger operators
or those with the power to compromise or compel access to the information on
the ledger.  For this reason, end users would not have possession of their own
CBDC funds, either.

Also, in line with the Federal Reserve paper, the Bank of England consultation
paper also states that users would face holding limits,  citing monetary and
financial ``stability'' and the likelihood of bank runs resulting from a
cascade of withdrawals as the justification for imposing limits upon the amount
of CBDC that individuals or businesses may possess~\cite{boe2023}, although
their possession of cash is not subject to similar limits.  However, protection
against bank runs involving CBDC can be realised by imposing restrictions on
withdrawals of CBDC from banks, just as it can be realised by imposing
restrictions on withdrawals of cash from banks.  Applying a limit to
withdrawals rather than holdings is more direct and appropriate mechanism to
safeguard against bank runs, since it is really the withdrawals by consumers
and businesses, not the total holdings by consumers and businesses, that
potentially introduce strain on the balance sheets of banks.  The choice to
impose the restriction on holdings rather than withdrawals is therefore not
justified, and an alternative possibility is more apt.  We note that
there are no limits to how much cash an individual can possess.
Were CBDC to become a \textit{de facto} replacement for cash in the future,
would an important right be lost in the process?

Finally, the consultation paper rejects the idea of bearer instruments, arguing
that ``a bearer instrument approach [...] would lead to completely anonymous
payments''~\cite{boe2023}.  The argument is similar to the argument made in the
consultation paper produced by the Federal Reserve, and it is similarly false.

\item\cz{Role of identity.}  The design described in the consultation paper
uses balances rather than tokens, and the implications for the rights of asset
owners are not explored or justified.  It is assumed that wallets would allow
users to ``[access] digital pounds, make payments, view balances and
transaction history''~\cite{boe2023}.  The functions of wallets as described
are not different than the functions of accounts, especially given the fact
that ``the private sector [...] would provide digital pass-through
wallets''~\cite{boe2023}, along with the implicit assumption that owners of
digital pounds would be required to identify themselves to PIPs to access their
own money.

\item\cz{Role of the ledger and scalability.} The CBDC design described in the
consultation paper makes use of a ``core ledger'' that records the assets held
by individuals.  Independently of the privacy considerations and the question
of tokens versus balances, the choice to rely upon the core ledger to record
assets in this way implies that all transactions must be recorded on the core
ledger.  This has two problematic consequences for the scalability of the
system.

The first consequence of relying upon the core ledger to store transactions
directly is that users must involve the core ledger in every transaction.
Because a transaction cannot be consummated until the core ledger is consulted,
the operators of the core ledger must be reachable both by the transacting
parties (and, if the ledger is distributed, also by each other) during the
course of every transaction.  Transacting parties must have network
connectivity to the core ledger, and they must wait for any system level delays
resulting from other transactions taking place at the same time.

The second consequence of relying upon the core ledger to store transactions
directly is that the ledger grows as a function of the number of transactions,
which implies that increasing the number of transactions will introduce stress
on the mechanism that maintains the core ledger, creating the opportunity for
denial of service by adversarial actors.  Cryptocurrency systems with this
design, such as Ethereum, typically address this vulnerability by introducing a
transaction charge (also known as a ``gas fee''), which introduces additional
economic frictions as well as fairness considerations related to the question
of whose transactions have precedence and whether there should be implicit
subsidy for users who might otherwise not be able to afford such charges.

A better design would not rely upon the representation of assets (tokens or
balances) directly on the core ledger (distributed or centralised).

\item\cz{Role of the issuer.} The CBDC design described in the consultation
paper specifies an outsized role for the central bank, which not only issues
tokens but also processes transactions, sets the rules, and maintains the
historical record.  It is specified that ``the wallet simply passes
instructions from the user to the [Bank's] core ledger'', indicating that all
assets are actually managed centrally, and that the ``core ledger would be a
single piece of infrastructure''~\cite{boe2023}.  This highly centralised
design is unlike any general-purpose payment system that has ever existed
before in the United Kingdom.  This design implies that the issuer is required
to process transactions directly, introducing a tremendous operational,
technical, and legal burden upon the central bank.

Because the system proposed in the consultation paper is highly centralised,
its users are exposed to system failure on the part of the central actor, and
the central actor is a high-value target for attackers as well.

\end{enumerate}

In March 2022, the Bank of England announced Project Rosalind, a partnership
with the Bank for International Settlements (BIS) for designing an API
prototype for retail CBDC based upon the architectural insights articulated in
its prior consultations~\cite{bis2022a}.  In October 2022, the Bank of England
launched a two-week procurement calling for private-sector businesses that had
previously registered with its Digital Marketplace to apply to develop a
proof-of-concept implementation of CBDC and to undertake research on offline
payments~\cite{uk2022}.  Then, in June 2023, two weeks prior to the extended
closing date of the 2023 ``digital pound'' consultation, the BIS Innovation Hub
revealed the names of the Project Rosalind partners, including the API users,
which included Mastercard, Amazon, and IDEMIA, as well as the advisers, which
included Stripe, Google, and Visa~\cite{othp69}, all of whom are active
participants in the incumbent identity-linked digital payments system based on
custodial accounts.

On 4 July 2023, less than one week after the close of the consultation, the
Bank of England updated the Terms of Reference for its CBDC Technology Forum,
including a new call for participation and an action plan to establish
subgroups to begin design work based upon the specific architecture that it had
described in its consultation paper~\cite{boe2023a}.  Six months later, in
January 2024, the Bank of England published its response to the consultation,
defending the positions that it had articulated in its proposal, once again
leaving them broadly intact~\cite{boe2024}.  The refusal to consider changes to
the design was striking, particularly given the manifest concern articulated by
the respondents, especially with respect to the rights of users of money in
general and privacy in particular.

\subsection{The Eurozone: the European Central Bank}

In June 2023, the European Commission (EC) published a proposal for a new
European regulation on the establishment of a digital euro~\cite{ec2023} along
with a press release describing a ``Single Currency Package [...] to support
the use of cash and propose a framework for a digital euro''~\cite{ecpr2023}.

Similarly to other CBDC proposals, the specification of the rules for the
digital euro in proposed regulation implies a specific set of policy outcomes:

\begin{enumerate}

\item\cz{Privacy.} The proposal specifically expresses support for transactions
with a similar level of privacy as cash, specifying that this degree of privacy
should apply to ``offline'' payments, whereas ``online'' payments should have a
similar degree of privacy to the electronic payments that are facilitated by
banks and card networks today.

However, given that retail payments in the EU are increasingly online rather
than offline, with the value transacted electronically now exceeding \euro{240}
trillion per year~\cite{ec2023a}, it is fair to say that the need for privacy
is precisely about payments in which at least one party is online, and the
proposal by the EC suggests that for transactions with the digital euro that
are not ``fully-offline'', that is, in which neither party is connected to a
network beyond each other, the degree of privacy will match the degree of
privacy offered by prevailing electronic transactions facilitated by banks and
credit providers, such as card payments.  Unfortunately, the privacy offered by
such institutions is actually a limited form of confidentiality, or
\textit{privacy by promise}, as those institutions have access to information
specifying not only how consumers spend their money, but also metadata, such as
location and device information, when they do.  It is fair to say that
consumers are subject to profiling and surveillance whenever they spend their
money in the prevailing electronic transactions referenced by the EC proposal,
so to say that consumers would have the same degree of privacy as they would
have in such transactions is to say that consumers would not have privacy when
they use the digital euro.

\item\cz{Motivation.} The arguments favouring the protection of cash acceptance
and access described in the Single Currency Package are clear: Persons in the
Eurozone have a right to use cash, and this right implies both the right to
obtain cash (``access'' to physical cash, for example from bank branches, ATMs,
post offices, and so on) and the right to use cash in their retail payments
(``acceptance'' of physical cash by retail merchants, in contrast to a more
limited concept of legal tender that refers only to the settlement of debts).
So, we can conclude that the EC intends to protect cash as a means of payment.
However, the attention given to ``fully-offline'' payments is curious in this
context, since the motivation for CBDC is not really about the unsuitability of
cash for offline transactions going forward, but instead is precisely about the
well-established decline of cash in favour of transactions in which one party
is online~\cite{access,tischer}, specifically both e-commerce via the Internet
and card and e-money payments at the point of sale.  It is not a stretch,
therefore, to say that the focus on offline payments is misdirected effort at
best, perhaps even disingenuous, leaving the question of the real intention and
of which stakeholders are responsible.

\item\cz{Certified hardware for ``fully-offline'' payments.}  A careful reading
shows that the reference to ``offline'' payments in the proposal for the
digital euro really refers only to ``fully-offline'' transactions wherein both
parties are completely disconnected from any third parties.  It has been
theoretically established that fair exchange between two parties is impossible
without the involvement of a trusted third party~\cite{pagnia1999}, so in
practice, the assumption that ``fully-offline'' payments are possible would
necessitate the involvement of a third-party actor at the point of sale.  This
is ostensibly achievable if at least one party to the transaction has a device
that carries out the will of a third party in preference to the will of either
of the transacting parties, although in practice that implies the use of
so-called ``trusted computing'': a ``trusted execution environment'' or
``secure element'' that can attest to a third party that the device had or had
not taken a certain action~\cite{tcg2004,trusted-computing}.  This kind of
remote attestation has been routinely criticised by technical experts and
others who advocate for the rights of users to exercise control over their own
devices~\cite{abelson2021}.  We have previously argued that it would be
unreasonable to impose certified hardware requirements on non-custodial
wallets~\cite{goodell2022a}:

\begin{enumerate}

\item As a basis for the security of a digital money architecture, the security
model of trusted computing introduces systemic risk,

\item The inability for a user to analyse the behaviour of a device undermines
trust,

\item The inability of experts from the public to produce compatible devices on
their own discourages innovation,

\item The special privilege granted to manufacturers of trusted devices is
incompatible with market-based competition, and

\item The trusted hardware constitutes a \textit{de facto} custodian, with a
relationship that is likely to carry high switching costs.

\end{enumerate}

Instead, hardware wallets should focus mainly upon transactions in which a
third party is externally reachable, or in which a third party has previously
facilitated the transfer of control without possession~\cite{goodell2022}.

\item\cz{Use of accounts.}  Concordantly, the EC proposal makes many references
to ``digital euro payment accounts''~\cite{ec2023}, implicitly suggesting that
this would be the primary mechanism by which individual persons would access
the digital euro.  By suggesting that the digital euro is to be ``available to
natural and legal persons''~\cite{ec2023}, what the EC proposal seems to
suggest is that service providers would offer a new kind of account that would
use the digital euro on the back-end to facilitate payments, and that the
digital euro would not be held directly by individual owners.  And so, the
architecture of interbank and card-based payment systems, along with the
business model enabled by that architecture, would be preserved in preference
to the architecture and business models associated with cash-based payment
systems.  Providers of ``digital euro payment accounts'' would have the
technical ability to monitor and restrict how a user spends his or her money,
just as providers of bank accounts or credit facilities do today.

\item\cz{Custody.} Even whilst such custodial services might be practical in
some cases, self-custody is prerequisite for both privacy and genuine end-user
ownership of money.  ``Wallet'' services provided by third-parties are no
substitute for non-custodial wallets in which individuals possess and control
their own assets.  Additionally, the suggested integration with the European
Digital Identity Wallet also introduces the possibility that the assets that a
natural person owns and how that person spends them would be directly
associated with the singular, physical identity of that person.  Finally, like
the Bank of England and HM Treasury in the UK, the EC proposal also envisions
implementing holding limits on the wallets of individuals, rather than
considering other ways to prevent individuals from acquiring CBDC too quickly.
Considered collectively, the characteristics envisioned by the EC for how
digital euro assets are held, if implemented, would have a chilling and
manipulative effect on how people spend their money, raising the question of
whether their money actually belongs to them or is merely licensed to them on a
contingent basis.

\end{enumerate}

The EC proposal for the digital euro does not explain the role that
``fully-offline'' payments would play in the future economy.  If it is assumed
that preserving cash is important, then why should the proposal focus so much
on ``fully-offline'' payments, when the existing cash system is both practical
and sufficient for this use case?  Furthermore, what is the basis for assuming
that the privacy of ``fully-offline'' payments is the only privacy by design
worth protecting, when online payments are substituting for cash in practice
and are increasingly important to everyday life in Europe?  Research suggests
that an economy that supports both CBDC and cash might have lower welfare than
an economy that supports either form of money alone~\cite{davoodalhosseini2022}
and that the deployment of a cash-like CBDC that competes with cash might make
using cash less efficient~\cite{agur2022}.  If this turns out to be the case,
and ``fully-offline'' CBDC turns out to be successful, then it is difficult to
imagine how the ECB will be able to maintain its headline support of cash for
long.  More importantly, by focussing the deployment of cash-like CBDC in
precisely the environments in which it will compete with cash rather than with
custodial payments, the ECB will have missed an opportunity to preserve the
most beneficial features of cash, with their concomitant benefits for privacy
and self-custody, in the burgeoning digital economy.  Instead, the effect would
be to bury what remains of those benefits.

\begin{table}[ht]

\begin{center}
\sf\begin{tabular}{l>{\raggedright}p{3.5cm}>{\raggedright}p{3.5cm}>{\raggedright\arraybackslash}p{3.5cm}}\toprule
Thematic element & Digital dollar & Digital pound & Digital euro \\\midrule
Stated motivation & improved efficiency and convenience for retail payments & economic growth, digital transformation & financial stability, monetary sovereignty, strategic autonomy \\\midrule
Use of accounts & users access CBDC via accounts with intermediaries & PIPs provide ``wallets'' that are \textit{de facto} accounts & service providers offer ``digital euro payment accounts'' \\\midrule
Privacy & privacy is subordinate to a transparency requirement; payers can be de-anonymised & payers can be de-anonymised & no privacy for ``online'' payments; limited confidentiality for ``offline'' payments inside the security envelope of a service provider \\\midrule
Custody & service providers hold funds in accounts & wallets operated by PIPs hold funds & service providers hold funds in accounts \\\midrule
Role of identity & services used to access accounts would be ``identity verified'' & an ``alias service'' links pseudonyms to a verified identity & ``digital euro payment accounts'' are subject to identification requirements \\\midrule
Holding limits & holding limits are applied by intermediaries on a per-user basis & holding limits are applied to CBDC wallets & holding limits are applied to digital euro payment acounts \\
\bottomrule\end{tabular}\rm
\end{center}

\caption{Common themes observed among the CBDC proposals.}
\label{t:assumptions}

\end{table}

\section{Themes}
\label{s:themes}

As can be seen from the well-publicised digital currency proposals around the
world (see Table~\ref{t:assumptions}), some common assumptions have begun to
materialise, prompting us to seek an explanation.  We have observed that
incumbent businesses have joined CBDC research programmes orchestrated by
central banks, thus seizing the opportunity to shape the narrative toward their
existing business models.  In addition, trans-national regulatory
organisations, such as the Financial Action Task Force (FATF), have reacted
unfavourably to non-custodial and privacy-enhancing electronic payment systems.
Meanwhile, central banks have used forums, such as the BIS, to share ideas
about the design of CBDC systems without much public debate, thus creating an
echo chamber.  Whether the common assumptions about the design of CBDC result
from a combination of these or other geopolitical factors, their similarities
are no coincidence.

\subsection{Asset ownership}

The report delivered by the Federal Reserve suggests that the litmus test for
CBDC should be its added value over existing payment mechanisms, implicitly
assuming that existing payment mechanisms will continue to exist and be useful
in all relevant contexts.  The report articulates a set of assumptions about
the design of CBDC and argues that CBDC must offer specific benefits or
improvements over existing methods of payment, while offering only a muted
acknowledgement of the decline of the use of cash in retail
transactions~\cite{fed2022}.  The report reiterates the US commitment to
``ensuring the continued safety and availability of cash'' but completely
ignores the costs of maintaining cash infrastructure.  Such costs have been
acknowledged as important by other central banks around the world, including
the People's Bank of China (PBOC), which cited the high fixed costs of cash as
a primary motivation for CBDC~\cite{pboc2021} as well as the central bank of
Sweden, which has acknowledged that for cash to survive, it must be explicitly
protected~\cite{riksbank2020}.  The Reserve Bank of New Zealand echoed this
observation, arguing that money requires special ``stewardship'' that had not
been required in the past~\cite{nz2021}.  If the decline of cash is indeed a
driver for the development of retail CBDC, whether for cost or consumer
behaviour or other reasons, it is because the properties and value add that
cash promotes are worth protecting. Were this not the case, then there would be
no value in protecting cash whatsoever.

A report commissioned by the United States Congress and delivered by the
Congressional Research Service (CRS) similarly ignores considerations of
cash-like properties, which include the ability of a user of cash to directly
possess and control it~\cite{crs2022}.  In its description of the differences
between CBDC and the current system, the report seems to suggest that the
primary weaknesses of the current system are counterparty risk and costs to
users of money, such as personal convenience and system-level
efficiency~\cite{crs2022}, and that those weaknesses constitute the main
argument for CBDC.  The report also asks where individuals ``would be permitted
to store and access CBDC''~\cite{crs2022}, suggesting that the choice would be
between financial institutions and the central bank itself, thus setting out a
straw man of only two options, as though no other options existed.  The report
completely ignores the possibility of non-custodial or ``unhosted'' wallets,
which the US Financial Crimes Enforcement Network (FinCEN) had explicitly
acknowledged as important in its 2020 consultation.  FinCEN even suggested the
possibility of establishing specific reporting requirements for transactions
involving such wallets~\cite{fincen2020}.  The CRS report does not consider
non-custodial wallets, omitting a salient motivation for digital currency in
its assessment of all wallets as not fundamentally different from accounts.

These examples highlight, at first instance, the framing of the debate.
Instead of taking a holistic approach to the development of money
and payments across each of its mediums, it instead looks exclusively
at the digital, comparing and contrasting existing electronic
retail payment mechanisms that are wholly mediated by custodial accounts
with possible CBDC designs, rather than, and more importantly, the
much broader landscape of methods and modalities of payment and payment
infrastructure, which includes, at its centre, the use of cash and
cash-like instruments.

\subsection{Privacy and criminal activity}

The Federal Reserve report states that privacy and deterrence of criminal
activity are fundamentally conflicting values: ``Any CBDC would need to strike
an appropriate balance, however, between safeguarding the privacy rights of
consumers and affording the transparency necessary to deter criminal
activity''~\cite{fed2022}.  The statement that these values are in conflict
exposes an assumption that the two desiderata cannot both be satisfied through
appropriate design, and it also raises the question of who would adjudicate the
``balance''.  The report states that CBDC intermediaries must verify the
identities of all persons ``accessing'' CBDC, for the purpose of providing
transparency for law enforcement and regulatory compliance~\cite{fed2022},
similarly to how financial intermediaries maintain records that identify the
sender and recipient of transactions between bank accounts.  A report published
by the UK House of Lords, following a consultation, also assumes that CBDC must
track consumer spending behaviour in the same manner as current electronic
retail payment mechanisms, which are mediated by custodial accounts, and
further suggests that concerns about the lack of consumer privacy might be an
inexorable aspect of CBDC in general~\cite{lords2022}.

However, and contrary to the presumptions of each of these two reports, digital
currency differs from bank accounts precisely because it can be stored in
unhosted wallets outside custodial relationships, a longstanding feature of
cash payments.  If the requirement for identity verification is construed to
imply that every transaction must be linked to the identities of both of its
counterparties, either via the prevailing model for custodial bank accounts or
via wallet identification, then consumers would not have privacy when they use
CBDC.  The assumption that regulatory compliance and crime prevention imply the
ability to compromise the privacy of consumers by tracing all payments to their
payers is not justified.  Although the Federal Reserve report does not state
this assumption explicitly, it is implied by the idea that intermediaries could
leverage existing tools for managing electronic transfers, wherein all assets
are held by custodians~\cite{fed2022}.

The House of Lords report states that ``[w]idespread adoption of any CBDC would
depend upon a high level of public trust'' concerning privacy~\cite{lords2022}.
We suggest that the ability for consumers to retain the ability to engage with
the digital economy without exposing their transactions to the risk of
profiling should be a primary motivation for CBDC.  Various architectures that
require identification of recipients without linking transactions to payers
have been proposed~\cite{chaum2021,goodell2022}, and with non-custodial
wallets, it is technically possible to verify the identities of consumers who
receive CBDC without allowing them to be linked to their spending habits.
Thus, consumer privacy and compliance objectives can be achieved
simultaneously, without relying upon ``data protection'' by a trusted third
party and thereby driving a decentralised trust at the centre of digital
payments.  By technically limiting the amount of data about consumer
transactions that are ever received by third parties, privacy by design
significantly mitigates the requirement for public trust.

The G7 report from 2021 goes still further than the House of Lords report and
states that privacy is in conflict not only with fighting crime but also with
the business models upon which CBDC might be designed to depend, stating that
privacy by design could ``reduce the range of possible business models in a
CBDC system''~\cite{g7}.  The interest in protecting ``business models''
references the possibility that some stakeholders, such as financial
intermediaries or platform service providers, might have interests that are
precisely at odds with consumer privacy, specifically in harvesting consumer
data for business purposes, and that the willingness of such stakeholders to
cooperate should be considered a precondition for deploying new infrastructure
for retail payments.  It seems that maintaining consumer privacy, something
that, in the realm of payments, has been enjoyed through the use of cash, is at
odds with private sector commercial interests, which wish to track the spending
patterns, location, and habits of consumers.  The G7 report seems to imply that
privacy ought therefore to be curtailed by to protect those private sector
commercial interests. This is clearly at odds with the interests of users and
consumers.

\subsection{Banking versus payment services}

Several of the CBDC proposals express concern about disintermediation, the
systematic withdrawal of consumer funds from bank accounts into CBDC wallets,
as a risk to financial stability.  Specifically, they reference the possibility
that CBDC would contribute to bank runs in ways that cash would
not~\cite{lords2022}.  The question of whether CBDC is really a substitute for
bank deposits arises from the question of why retail consumers use bank
accounts in the first instance.  CBDC can be designed so that it is useful
primarily as a medium of exchange for making retail payments, rather than a
long-term store of value~\cite{barrdear2022}.  If designed in this manner, it
is not really much different from cash and would not necessarily compete with
bank deposits as a store of value.  Specifically:

\begin{itemize}

\item\cz{Not charging interest.} It is possible to design CBDC so that, like
cash, it does not earn interest.  An article published by the IMF argued that
the extent to which CBDC competes with bank deposits is largely determined by
whether it pays interest~\cite{adrian2019}, a view echoed by the CRS
report~\cite{crs2022}.  Operationally speaking, it would be natural to assume
that CBDC assets would not be rehypothecated, as in most cases it is not
possible for a central bank to make risky investments~\cite{fernandez2020}.  In
any event, if CBDC does not earn interest, or even if it earns interest but
yields less than commercial bank deposits, justification for the argument that
consumers would actively prefer CBDC to bank deposits is unclear.

\item\cz{Limiting withdrawals.} It is also possible to design CBDC so that
withdrawals of CBDC from financial institutions would be operationally
restricted in a manner similar to how withdrawals of cash from financial
institutions are restricted.  Subject to those design characteristics, CBDC
would not present a significant risk of large-scale outflow of funds from
consumer bank accounts, both because CBDC is an inferior substitute for bank
deposits as a store of value over long periods of time and also because
withdrawals would be systematically limited.  We also note that nation states
have imposed limits to bank withdrawals during periods of crisis.  For example,
in 2015, Greece prevented account holders from withdrawing more than 60 euros
per day to protect its banks as its government negotiated the terms of its
bailout~\cite{apnews2018}.

\end{itemize}

With its digital euro proposal, the European Commission explicitly recognised
``safeguarding financial stability and financial intermediation'' as an
objective that should be addressed by limiting the ``excessive use of the
digital euro as a store of value''~\cite{ec2023}.  The European Commission also
clearly stated that the primary use of the digital euro would be as a means of
payment, and that to achieve this objective, it ``should not bear
interest''~\cite{ec2023}.  Similarly, the House of Lords report states that the
two ``main options'' for ameliorating the risk of disintermediation is to
``disincentivise use'' through unattractive interest rates for retail consumers
or to ``limit the amount of CBDC that can be held or spent''~\cite{lords2022}.
There is no particular reason to assume that CBDC must pay interest.  In fact,
if CBDC were designed to be cash-like, then we might assume that it does not
pay interest at all, at least in the first instance.  Furthermore,
notwithstanding the recommendations of some of the CBDC proposals, limits to
the amount of CBDC held by consumers can be achieved indirectly, via withdrawal
limits and expiration policies instead, and does not require imposing holding
limits or other constraints on non-custodial wallets used by consumers, or any
other mechanism that would force consumers into unwanted relationships and
necessitate revealing or utilising personally identifiable information when
they transact.

We conclude that these proposals and reports implicitly assume that consumers
value bank accounts primarily as a means for making electronic payments.  We
question this assumption, although we note that payment services, which
generate revenue from data harvesting and transaction fees, constitute a
significant source of income for some financial services
businesses~\cite{radecki1999}.  If it turns out to be the case that revenue
from such services has become essential to the profitability of commercial
banks that serve consumers, then we might reasonably conclude that the role of
consumer-facing banks has potentially shifted with the decline of cash.  In
other words, banks have diversified their core business activities away from
banking.  Such a conclusion would raise the question of whether consumer-facing
banks continue to be fit for purpose, if a significant source of their income
derives from data harvesting and transaction fees as opposed to interest on
lending.

\section{Institutional motivations for retail CBDC}
\label{s:motivations}

Governments and businesses have supported and promoted the development of
retail CBDC for a variety of reasons.  Chief among them are preservation of
monetary and financial sovereignty, protection of existing business models, and
institutional security against misbehaviour on the part of users of money.  It
is fair to say that these motivations are largely reactionary in nature.  This
section explores these motivations.

\subsection{Monetary and financial sovereignty}

An important driving force underpinning the pursuit of CBDC involves
institutional concerns about emerging risks to monetary and financial
sovereignty.  Unsurprisingly, such concerns are of particular interest to
central banks and governments, who are observing a rapidly changing payments
landscape.  As retail consumers face a plethora of new ways to make payments,
it has become clear that inaction on digital currency is not a particularly
safe choice.  The set of risks comprises two broad categories: the availability
of mechanisms for consumers to make payments with \textit{private money}, by
which we mean cryptocurrencies, privately issued tokens and stablecoins, and
e-money in closed-loop payment systems, and the availability of mechanisms for
consumers to make payments with \textit{foreign money}, the currency issued by
the central banks of other countries.  Both scenarios can undermine the ability
of central banks to implement monetary policy and can impact the monetary and
financial sovereignty of a nation and its sovereign tender.

\subsubsection{Private money}

In recent years, several innovations have highlighted the risks of private
money creation.  Although these innovations are in varying stages of
realisation, private-sector actors have potentially tremendous incentives to
pursue them:

\begin{itemize}

\item \cz{Closed-loop payment systems.}  According to its 2021 report, the PBOC
considers mobile payment systems to be a primary motivation for the development
of CBDC~\cite{pboc2021}, and the reason is clear.  Platform service providers
such as Alibaba and Tencent have developed payment systems that allow users to
make payments within a network of platform users, creating a closed
loop~\cite{zhu2017}.  Open-loop payment systems, which are popular in most of
the world outside China, are effectively fancy wrappers around traditional
banking infrastructure, focusing mostly on user experience and convenience
factors.  Open-loop payments are debited from one bank account and credited to
another.  In contrast, closed-loop systems allow transactions between platform
users without the involvement of regulated financial intermediaries or
traditional payment networks~\cite{lai2020}.  The platform operator generally
has a commingled bank account for its users, and the recipient of a payment
generally does not receive the funds into a bank account in his or her name.
Instead, the platform operator keeps track of the allocation of funds among its
clients and users.

Closed-loop payment systems, such as Alipay and WeChat Pay, have been
criticised by the government of China, which adopted ``a series of measures to
slow the tech companies, enhance the government's role, and possibly bring
payments back into a bank-centric system''~\cite{klein2021}.  Government
regulators might legitimately fear losing the ability to monitor economic
transfers that should be eligible for taxation or compliance checks.  In the
absence of regulation, closed-loop payment platforms have not only the ability
to collect data for profit, but also the ability to withhold such data from
authorities.  It is plain to see why, following the adoption of closed-loop
payment platforms by ordinary users as a general-purpose method for remitting
payments, the government of China established rules to rein in operators of
those platforms~\cite{knight2021}.

\item \cz{Corporate digital currency.} Corporations have a variety of reasons
for wanting to facilitate payments using money of their own creation, and the
mechanisms deployed for such purposes take a variety of forms.  Customer
loyalty programmes, such as those offered by airlines or coffee shops, allow
consumers to accumulate points in exchange for their patronage, and the points
can be redeemed for goods or services with partners participating in the
loyalty programmes.  In some ways, such programmes resemble closed-loop payment
systems, although points are generally useable only for consumption and are not
exchangeable for fiat currency.  Such programmes can be an important source of
revenue for their sponsors, particularly if the sponsor sells the right to
participate, invests the proceeds, and buys protection against the
risk~\cite{bloomberg2017}.

In recent years, businesses have explored the possibility of digital currency
as a kind of closed-loop payment system on a global scale.  In particular,
excitement about digital currency projects by Facebook/Meta (Libra, Diem, Zuck
Bucks, and so on) and by payment networks such as
MasterCard~\cite{mastercard2021,mason2021} suggest the possibility that
consumers and merchants could dispense with fiat currencies entirely, placing
their faith in private issuers instead, with, for example, multi-currency
stablecoins.

\item \cz{Stablecoins and their analogues.} We use the term \textit{stablecoin}
to denote a privately-issued token whose value is pegged to the value of
another asset, such as a currency issued by a central bank or a commodity
asset.  The peg is generally maintained by the issuer.  Stablecoins can be
further divided into two categories: \textit{algorithmic stablecoins}, whose
value is maintained by trading activity (including issuance and redemption) and
\textit{collateralised stablecoins}, which are backed by assets held by the
issuer.  Stablecoins backed by central-bank reserves have also been
proposed~\cite{goel2023}.

From the perspective of risk, stablecoins can be compared to assets traded in
money markets such as bankers' acceptances, commercial paper, and so on.
However, unlike those assets, stablecoins can be traded via a digital
infrastructure outside of regulatory oversight.  The ability of ordinary
persons in some country to conduct commerce using stablecoins pegged to
something other than the currency of that country might pose a threat to
monetary sovereignty, particularly for central banks and governments with poor
credit or unstable currency~\cite{gorton2022}.  This concern was specifically
recognised when Facebook (now Meta) first introduced the now-defunct Libra
project~\cite{tang2019}.  Meanwhile, regulators of countries with strong
currencies have expressed concern about the business model underpinning
stablecoins, suggesting the possibility of a tussle in the medium
term~\cite{panetta2022}.

\end{itemize}

\subsubsection{Foreign money}

Some arguments supporting the adoption of digital currency focus upon the use
of foreign money in domestic contexts and the risk associated with state actors
more generally.  These risks are about not only the adoption of central bank
digital currency, but the capture of value associated with the digital economy
more generally.

\begin{itemize}

\item \cz{Currency substitution.}  Some governments have expressed concern that
domestic consumers already prefer foreign currency.  Specifically, reduced
barriers to international trade concomitant with the rise of the digital
economy may favour the proliferation and use of foreign currency in preference
to domestic currency in some countries, particularly governments whose national
currencies are free-floating but whose condition is such that they cannot avoid
secular inflation.  For example, consider El Salvador, whose leaders opted to
accept Bitcoin as legal tender~\cite{renteria2021}, ostensibly following
pressure to take action to avoid perceived ``dollarisation''.

\item \cz{Regulations to manage the ``spillover'' risk of foreign CBDC.} Access
to the CBDC of foreign countries has the potential to ``lead to currency
substitution and loss of monetary sovereignty in both the issuing and foreign
country, which in turn might impede the ability of authorities to achieve their
own policy objectives, including monetary and financial stability, and the
countering of illicit finance'' \cite{g7}.  An economic model developed by
researchers at the European Central Bank suggests that the use of foreign CBDC
could amplify the transmission of shocks and increase systemic
risk~\cite{ferrari2022}.  Arguably, part of the rush by governments to design
CBDC systems is related to the impetus to ensure that they will have a voice in
the creation of regulatory mechanisms and standards that will underpin the
global rules for CBDC, and that they will be ready to protect themselves from
the effects of foreign CBDC in the event that it is deployed.

One can easily imagine an outcome in which tourists visiting a particular
destination opt to use the CBDC of their home country, and that, through
platforms of payment service providers, it is easily offered and accepted by
merchants in the destination country.  Were this to become widespread across
tourism, trade, and other areas, it could impact a nation's monetary and
financial sovereignty.

\item \cz{Opportunity to create a common currency.}  The flip side of the
currency substitution risk is an opportunity for trans-national communities
with strong economic ties to create a similarly strong common currency for
trade.  Whether realistic or not, digital currency could be seen as an
inexpensive and politically feasible way to introduce a mechanism for
conducting trade without incurring the costs associated with establishing a
banking system in the new currency, managing a new physical currency
infrastructure, or negotiating new rules with neighbours.

\item \cz{The ``Sputnik effect''.}  Some governments perceive a race to unlock
economic potential from digital currency, fuelled by the perception of
opportunity by private-sector businesses to capitalise on the potential for
digital transformation.  This race has geopolitical importance, not only
because governments can potentially use CBDC to promote innovation by their
domestic businesses, but also because their domestic businesses might be able
to capture value from new, international value chains associated with
technologies and services that will underpin new mechanisms for digital
payments.

Suppose, for example, that a particular nation builds out a suite of digital
tools that it and the businesses within its jurisdiction expect to export to
businesses and states of other nations.  Part of the provision will include the
innovations and intellectual property of that nation and its businesses.
Possibly, their own CBDC would be linked to it, and possibly not.  In either
case, however, that nation will be positioned to support and therefore
influence the value chain in the digital economies of other nations, towards
capturing some of its value.

\end{itemize}

\subsection{Lobbying opportunities for private-sector incumbents}

Much of the opportunity for private-sector businesses is related to the
potential to extract economic rent, for example, by providing identity services
to consumers or by securing the institutional acceptance of digital currency
standards that support their business models.  These opportunities are shared
between incumbent actors in financial services and the technology sector,
including both technology platform operators and the innovation ecosystems that
they lead.

\begin{itemize}

\item\cz{Banks and financial market participants.} Over the past several
decades, there has been a secular shift toward payments as a significant source
of revenue for consumer-facing banks~\cite{radecki1999}.  Possibly, this is
related to the downward trajectory of interest rates over the same
period~\cite{fred-dgs10} and the fact that low interest rates can reduce the
margins that consumer-facing banks earn from deposits, thus impacting their
profitability~\cite{ulate2021}.  We surmise that commercial banks were
incentivised to develop innovative solutions to maintain profitability in their
consumer-facing business lines.  Although interest rates have risen in response
to inflation in recent years, this new development is relatively recent, and
the value chains from payments-oriented businesses and infrastructure have
remained.  Global payments revenues have grown steadily over the longer period,
reaching 1.5 trillion dollars per annum, and analysts are projecting that this
growth will continue over the next decade~\cite{bcg2022}.  The possibility that
CBDC might substitute for bank deposits is seen as a risk factor with
implications for financial stability~\cite{kumhof2021}.  Although CBDC can be
designed to avoid being used as a long-term store of value~\cite{barrdear2022},
it is worth considering whether retail consumers establish bank accounts not
primarily as a store of value, as banks have traditionally been used by
depositors, but instead as a medium of exchange, a relatively new role of banks
as payment service providers first and foremost.  Concordantly, it might also
be worth considering whether the changing business model of consumer-facing
banks introduces a new example of systemic risk.

\item\cz{Technology service providers.} Banks and prevailing platforms for
electronic payments also face a reckoning as new technology platforms offer
significant competition through better user experience, cheaper fees, ease of
access, cross-border interoperability, and better interoperability across the
various ecosystems within existing platforms.  Consider the rise of Alipay and
WeChat Pay in China~\cite{klein2020}, which provide a wealth of high-assurance
personal data that are used to build detailed profiles of their users, which in
turn form the basis for credit decisions about those users~\cite{allen2022}.
For these reasons, social media and e-commerce platform operators are keen to
replicate their success in the West~\cite{hoskins2023}.  Certainly there is no
particular reason that banks must be the ones to provide payments
infrastructure; after all, payments infrastructure is more about
telecommunications than banking.  Specifically, it is more about the underlying
technological infrastructure and operating system and associated standards than
it is about being able to accept deposits and offer credit.

\item\cz{Data harvesters (a.k.a. ``Big Tech'').}  As we have learned from the
rise of surveillance capitalism, there is tremendous value in profiling
individuals~\cite{zuboff2019}.  As it turns out, payments are excellent
channels for accumulating data about individuals.  Because of AML regulations
demanding strong customer authentication as a way to deter criminal activity,
they offer links to high-assurance identity information.  Because the
transactions themselves cost money, mistakes and security breaches are costly,
and as a result, the data quality is higher than it otherwise would be.  And
finally, because nearly everyone must inexorably interact with the economy,
consumers cannot avoid making payments.  As e-commerce (that is, payments
conducted via the Internet) transactions supplant services traditionally
supplied by brick-and-mortar shops, and as merchants increasingly refuse cash
at the point of sale, electronic payments have captured an increasing share of
all retail payments.  The data accumulated from consumer payments has been
perceived as a way to monitor credit and therefore reduce risks associated with
lending and insurance, and is sometimes considered, somewhat dubiously, to be a
vector for promoting financial inclusion.

\end{itemize}

The assignment of current payments industry incumbents to roles in the future
digital payments landscape has not yet been resolved.  All of the current
market participants face uncertainty, and for them, participation in the
framing the narrative for the future of money in general, and CBDC in
particular, is a way to manage risk.  At the same time, it is also worth noting
the conspicuous absence of some participants in the contemporary payments
industry, most notably those involved in operating the infrastructure that
forms the basis of the cash economy, among the global payments conferences and
public debates about the future of money.

This absence is a strategic mistake for the providers of cash services as well
as a coup for providers of services that support and encourage consumers to use
custodial accounts to conduct payments.  The cash economy has supply chains
that include the businesses that design and manufacture central bank notes and
coins, the providers of security technology for the anti-counterfeiting
features of physical cash, ATM networks, cash security and transport
businesses, manufacturers of cash registers, and merchant-facing banks that
take cash deposits, among other businesses.

If CBDC is about providing a public payment option in the form of digital cash,
then the development of CBDC could be viewed less as an extension of existing
digital payments systems, and more as an adaptation of physical means of
payment to the digital economy.  In this case, it would stand to reason that
businesses involved in the supply chains that support cash would be natural
partners and instigators of any plan to create digital cash.  However, it is
not clear that those businesses have had much of a voice in public
conversations about CBDC to date.

\subsection{Governments, institutional incentives, and the Panopticon}

Institutions, including governments, have incentives to support the increased
collection of data about individuals.  Some of these incentives are related to
security practices, which can be understood in terms of the classic tussle
between security and privacy.  Other incentives are political, ranging from
interactions with international rule-making organisations to the perennial task
of justifying the use of taxpayer money.

\begin{itemize}

\item\cz{Crime and tax evasion.}  Some governments have associated the use of
cash with money laundering and corruption~\cite{morshed2021}, arguing that by
providing a universally-accepted digital payment option, CBDC is the key to
implementing a cashless society.  The governor of the central bank of
Bangladesh expressed a desire for all retail transactions to be settled by
digital or mobile technology by 2027, arguing that eliminating cash would
improve efficiency, improve financial inclusion, and reduce
crime~\cite{pesek2023}.  In this context, it is easy to understand why
government officials in Bangladesh seem to believe that if ``a company doesn't
accept digital payments, or if they strongly discourage it, that behaviour
itself should be considered suspect''~\cite{rhid2023}.  Similarly, Godwin
Emefiele, governor of the Central Bank of Nigeria, argued that a policy of
systematically eliminating the use of cash would facilitate the tracking of
funds~\cite{emejo2022}.  The e-Naira, a CBDC, was intended to facilitate the
transition to a cashless society, although it was ultimately rejected by the
public, which responded with protests~\cite{brody2023,shumba2023}.

\item\cz{Financial inclusion.}  Many CBDC proposals include designs that track
the spending habits of ordinary consumers.  The International Monetary Fund
published an argument citing CBDC as beneficial to financial inclusion because
``CBDC generates greater surplus in lending by reducing credit-risk information
asymmetry'' provided that ``is valuable to households as a means of payment or
for credit-building''~\cite{tan2023}.  The mechanism by which information
asymmetry is reduced is the ability of lenders to observe patterns in spending
activity among consumers.  A similar argument was previously made by the Bank
for International Settlements, which published an article stating that ``the
value of data generated by CBDC payments is also clear'' as a means of
supporting credit and insurance services to ``unserved or underserved segments
of the population''~\cite{auer2022}.  Some experts have argued that unlike
digital financial innovation offered by private-sector banking services, CBDC
can be promoted as a public payment option, and some experts have suggested
that CBDC can ``only promote financial inclusion and stability when they serve
public interest''~\cite{ozili2023}.  Similarly, the Asian Development Bank
argued that ``financial exclusion may be better addressed by the central banks
by using the data on individuals collected through their use of
CBDCs''~\cite{didenko2021}.

\item\cz{Institutional security.} Unsurprisingly, institutions and businesses
are motivated to collect intelligence to support their security procedures,
including both prevention and response.  Identity information provided with
payment transactions can be collected as part of a broader effort to institute
procedures for anomaly detection that can potentially aid in the prevention of
crime and terrorism that could threaten the well-functioning of institutions.

\end{itemize}

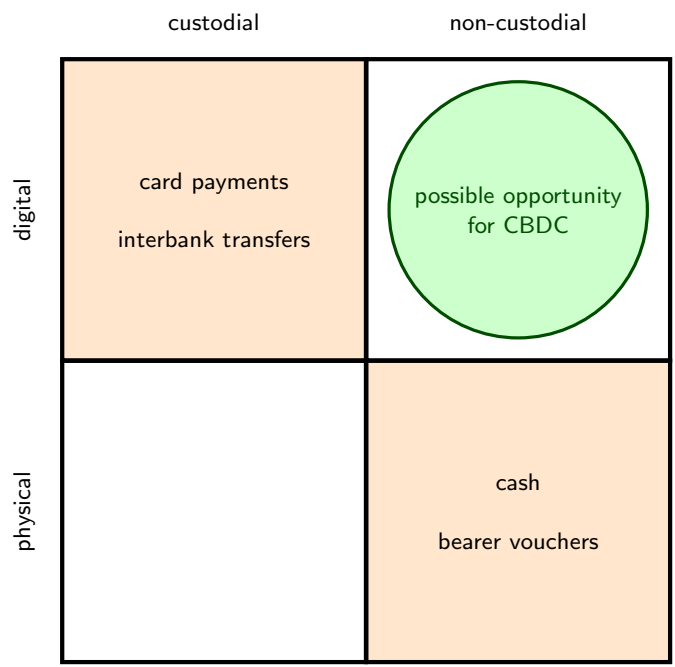
\begin{figure}[ht]

\def\sqw{10.7em}
\begin{center}
\hspace{-0.8em}\scalebox{1}{
\begin{tikzpicture}[>=latex, node distance=3cm, font={\sf \small}, auto]\ts
\tikzset{>={Latex[width=4mm,length=4mm]}}
\node (box1) at (-2,2) [box, text width=\sqw, text height=\sqw, fill=orange!20] {};
\node (box2) at (2,2) [box, text width=\sqw, text height=\sqw] {};
\node (box3) at (-2,-2) [box, text width=\sqw, text height=\sqw] {};
\node (box4) at (2,-2) [box, text width=\sqw, text height=\sqw, fill=orange!20] {};
\node (circ) at (2,2) [draw, circle, very thick, text width=9em, color=green!30!black, fill=green!20] {};
\node (n1) at (-2,4.5) {custodial};
\node (n2) at (2,4.5) {non-custodial};
\node[align=center] (t1) at (-2,2) {
    card payments\\
    \\
    interbank transfers
};  
\node[align=center, color=green!30!black] (t2) at (2,2) {
    possible opportunity\\for CBDC
};
\node[align=center] (t2) at (-2,-2) {};
\node[align=center] (t2) at (2,-2) {
    cash\\
    \\
    bearer vouchers
};  
\node[rotate=90] (n3) at (-4.5,2) {digital};
\node[rotate=90] (n4) at (-4.5,-2) {physical};
\end{tikzpicture}}
\end{center}

\caption{A representation of the universe of payment mechanisms.}
\label{f:money}

\end{figure}

\section{Accounts all the time}
\label{s:accounts}

Throughout the debate about the design of CBDC, the question of whether money
would be ``account-based'' or ``token-based'' has featured prominently.  Much
of this debate has centred around the assumption that token-based money, which
can be non-custodial, is either incompatible with a digital payment system or
incompatible with the existence or use of bank accounts.  This serves as a
false dichotomy between digital money and non-custodial money.  The use of
accounts is not a neutral design choice: Accounts intrinsically introduce at
least some degree of \textit{accountability} of the account-holder to the party
maintaining the account.  The real issue is about custody: It is possible for
money to be both digital and non-custodial (see Figure~\ref{f:money}).  There
are several lines of argumentation that might explain this misunderstanding:

\begin{itemize}

\item\cz{Lack of imagination.}  In the developed world, much of the public is
accustomed to an economy intermediated by smartphones, e-commerce platforms,
and card payments.  The architects of CBDC proposals are certainly familiar
with the reality that bank accounts underpin all of their interaction with
digital economy today, and it is conceivable that they have not imagined that
the future could be any different.  Services provided by third parties underpin
all interaction with the digital economy, and accounting relationships backed
by strong customer authentication are the way that AML/KYC requirements are
satisfied.  In this context, it might be reasonable to ask: \textit{If cash
were to be proposed as a means of payment today, then would it be considered
too risky to develop and deploy?}

\item\cz{Data access and control in the name of institutional security.}
Governments have an operational interest in collecting all possible data about
transactions, both \textit{ex ante}, for the purpose of detecting anomalous
behaviour, and \textit{ex post}, for the purpose of conducting investigations.
Data collection by governments is sometimes supported on the grounds that
surveillance facilitates the efficiency of investigators and the prevention of
crime and terrorism.  The justification for such collection can be couched in
terms of the paternalistic responsibility that an institution has to people in
its milieu.  Private-sector organisations and businesses sometimes use a
similar argument to support procedures that create profiles of their customers
or employees.  However, there are significant risks associated with large-scale
data collection, including the risk of collateral damage to institutions and
individuals alike resulting from data breaches in general, as well as the
cybersecurity risks associated with amassing large volumes of marketable
personal data in particular.

\item\cz{Geopolitical interests.}  Similarly, governments are keen to pursue
data collection in the interest of cooperation with international law
enforcement activities, and national governments also have an incentive to use
international rule-making organisations to project power beyond their borders.
Consider the recommendations of the FATF, and Recommendation 16 (the ``Travel
Rule'') in particular, which specifies that both the payer and recipient of
electronic transactions must generally be
discoverable~\cite{fatf-recommendations}.  A recommendation adopted in 1996
also states: ``Countries should further encourage in general the development of
modern and secure techniques of money management [...] as a means to encourage
the replacement of cash transfers''~\cite{fatf1996}.  In 2022, the FATF
published a recommendation explicitly stating that the Travel Rule would
require financial institutions ``to share relevant originator and beneficiary
information alongside virtual asset transactions'' and characterising
``unhosted wallets'' as an ``emerging risk''~\cite{fatf2022}.

The FATF recommendations have sparked controversy, particularly in recent
years.  Evidence suggests that the FATF has been used as a vehicle for
projecting geopolitical power, possibly in violation of the United Nations
Charter and the Vienna Convention of 1988~\cite{doyle2002}.  A report by the
Royal United Services Institute argued: ``Concrete governance does not underpin
the FATF’s decisions, even though they may substantially impact countries and
their economies. It can therefore be challenging to explain to government
ministers and officials how the FATF has the legitimacy necessary to make such
decisions''~\cite{rusi2024}.

In the European Union, there is an additional motivation to pursue CBDC as a
means to promote its continued development of its framework for electronic
identification, authentication, and trust services, eIDAS.  For example, in
2022, a member of the Executive Board of the Deutsche Bundesbank suggested that
the government-issued digital identity wallet proposed in the new eIDAS
framework (eIDAS 2.0) would be able to hold a digital euro~\cite{balz2022}.
The eIDAS 2.0 framework has faced public criticism for its implications for
personal privacy in the EU~\cite{hancock2022,biometric2023}.  In particular, a
group of prominent civil society organizations, academics, and research
institutions signed an open letter arguing that the implicit requirement for an
individual to use a government-issued wallet that inexorably links together his
or her transactions or credentials would ``undermine the privacy of EU
citizens'' and ``create an unprecedented risk for every European in their
online and offline life''~\cite{epicenter2023}.

\item\cz{Data subsidy.} Another concern is that governments and other providers
of infrastructure might prefer to work with businesses that collect and
monetise consumer data for profit.  In principle, profits from data revenues
can be used to offset the costs of building, deploying, and managing the
infrastructure.  As a result, public services are cheaper if they are run by
data harvesters, and infrastructure providers can justify the choice of such
vendors by arguing that saving money supports their duty of care to customers
or the public.  For example, Facebook (now Meta) offers a controversial service
called ``Free Basics'' to provide Internet services to users in developing
markets, without requiring them to pay money (specifically, working with local
carriers to ``zero-rate'' the services), but effectively in exchange for their
data~\cite{solon2017}.  Similarly, in November 2023, the data broker Palantir
won a bid to operate the data platform for the UK National Health
Service~\cite{armstrong2023}, affording it an opportunity to extract value from
vast volumes of personal information.  The implicit subsidy from data revenues
can apply at any stage of a project, and is evident from both the eagerness of
data analysis firms to engage with infrastructure data sources and the
competitiveness of their bids.

\item\cz{Cash as an inferior form of money.}  A line of argumentation has
emerged wherein money is cast as a representation of the history of the
behaviour of the actors in an economic system; this argument was championed
around the end of the last century by Narayana
Kocherlakota~\cite{kocherlakota1998}.  In this framing, cash is seen as an
imperfect substitute for credit-based approaches that maintain complete records
of the histories of individual actors, and the use of credit, in which payers
identify themselves, in preference to cash would yield greater economic
efficiency.  This argument is supported by the perspective that the main
benefit of cash is its use in criminal activity~\cite{camera2001} presumably
because, after all, parties to have nothing to hide would have nothing to fear
from using credit.  However, the assumption that privacy implies misbehaviour
is problematic in general~\cite{coustick2015}, and Charles Kahn and some
central bankers used a model to demonstrate that ``the value of [cash] money
may derive from its supposed imperfection, from the anonymity that it
confers''~\cite{kahn2005} and that removing cash would result in the loss of
trade.

\item\cz{Surveillance and the Panopticon effect.} Surveillance might be seen as
a way to control or address abusive behaviour despite the fact that doing so
might discourage legitimate but unorthodox patterns of behaviour.  Consider the
increasing use of technologies by employers to track individuals via their use
of personal devices~\cite{privacyinternational2022}, and the increasing use of
web security services, such as those provided by Akamai~\cite{akamai2023}, to
force customers of businesses that provide online services to reveal their IP
addresses, with the stated goal of incentivising good behaviour.  Mechanisms to
force customers to provide payment card details are no different.  Although it
is tempting to imagine that people are better-behaved when they are being
watched, it is important to consider that the original Panopticon, envisioned
as an efficient architecture of a prison with a central
tower~\cite{bentham1791}, was as much an instrument of punishment as a tool of
compliance~\cite{foucault1977}.  Since there is no way for ordinary individuals
to avoid interacting with the economy, do we really mean to create this kind of
infrastructure for the general public?  And before we assume that the
Panopticon would be effective, we must consider that sufficiently wealthy or
powerful individuals will find a means to coerce others to conduct transactions
at their behest, and potential criminals can expect to find a vast marketplace
for stolen identity credentials, including credit cards issued to victims who
satisfied AML/KYC criteria, at their disposal~\cite{ofcom2023}.

\item\cz{Maintaining the power to exercise unilateral control.} A final
possibility is couched in the twin mechanisms of surveillance and censorship.
By collecting volumes of personal data and related information about
transactions, coupled with building broad profiles of individual persons who
transact, service providers can more readily implement censorship of specific
individuals and circumstances to meet particular objectives.  Although the
means to determine who and when to censor is not the same as the power to be
able to censor a customer, consumer, or employee, the fact that customers are
locked into accounts maintained by specific third-party actors means that the
ability for customers to access and control their own assets, or to conduct
transfers or transactions, can be limited and restricted, at any time and for
any reason, by those third-party actors.  Since custodians (and therefore
governments) already enjoy such power over card and interbank payments,
relinquishing that power might require an act of great courage.

\end{itemize}

\section{Potentially missed opportunities}
\label{s:opportunities}

As nations grapple with the best CBDC design to develop and deploy over the
next several years, there is a reckoning concerning which nations will reap the
rewards of creating a viable digital payment infrastructure to underpin and
support the digital economy.  Nations adopting designs that, due to flaws
resulting from not prioritising the right properties and underlying principles,
may fall behind in the ultimate offering that is provided to users.
Ultimately, the categories are simple.  There are those that prioritise the
protection of business models or sovereignty-based interests in preference to
the interests of users, and there are those that treat public access to the
digital economy as part of the \textit{raison d'etre} of a central bank.  For
central banks in the latter category, those that incorporate properties of
cash, such as a token-based form of money that is accessible to all users,
privacy-preserving for consumers, non-discriminatory, and compatible with being
held and controlled in non-custodial wallets, will be able to reap rewards
across a variety of areas.

\subsection{Competition and Innovation}

One of the central tenets of market efficiency and the avoidance of monopolies,
oligopolies, and cartels is the requirement for competition among market
participants.  Competition would require payment service providers, as well as
providers of supporting services, to deliver the best service at the best price
for users.  If a market is sufficiently competitive, then organisations that
fail to be attentive to the interests, demands, and preferences of their users
and consumers would not survive for long.  Competition also implies the ability
for organisations to innovate and provide differentiated value propositions for
their customers and suppliers.

One central question is whether a particular direction for the design of
digital currency could be expected to nurture greater competition and
innovation than any other.  Were the models for digital currency based upon
custodial accounts to be adopted, then digital currency would be more akin to
bank deposits than to cash, and it would be likely that regulated financial
institutions, mostly consumer-facing banks, would be the only or primary
entities to provide services to users, even when there is no technical reason
for which this must be the case.  This approach can be expected to protect the
market structure for electronic payments of today, and we do not envisage any
major differences.  In contrast, were an approach to digital currency based
upon tokens and non-custodial wallets to be adopted instead, then digital
currency would be more akin to cash.  For the purpose of retail payments, not
only would banks, incumbent financial institutions, and money services
businesses continue to provide essential services, but other kinds of
businesses and organisations would have a place in the value chain as well,
including those that provide ancillary services for the purposes of
facilitating payments, providing hardware, maintaining software, offering
better user experiences, and even offering custody for digital assets, perhaps
coupled with security and management services.  The diversity in the range and
types of services, as well as the competition and innovation afforded by lower
barriers to entry, would enhance the efficiency and value delivered to users
and consumers.  Through the lens of competition, it is clear that designs
oriented around adapting cash for the digital age are preferable to those that
seek to build a new back-end for custodial business models.  An approach based
on tokens and non-custodial wallets would offer a greater diversity in the
kinds and types of organisations that can service the interface between
citizens and the digital economy.  It would also support a way for people to
choose between service providers on an as-needed basis, without being locked
into account-based relationships that can be used as an instrument of
surveillance and control, thus strengthening the potential for competition and
thus innovation in servicing retail payments in the economy.

\subsection{Privacy}

Considering the implications of the recent CBDC design proposals for end-user
privacy, one might assume that CBDC is inherently bad for privacy and the human
rights that depend upon privacy, but an important nuance remains.  The digital
economy has an increasingly important role in everyday life for an increasing
number of people around the world, and in this digital economy, consumers have
little or no real privacy.  Every time people tap a debit card at an EMV
terminal or enter bank account information into a web form to make a payment,
their locations, choices, and habits become part of their permanent record.
This is happening already, whether central banks deploy CBDC or not.

Institutionally-supported digital currency, including CBDC, offers an
opportunity for institutions to \textit{protect} the privacy of consumers, not
with ``privacy by promise'', but with real privacy by design using
privacy-enhancing technology.  Institutions and governments have every reason
to do so.  After all, accountability of consumers for their transaction history
introduces a form of friction that can undermine trade, and market efficiency
depends upon allowing consumers to make their own choices, free of the concern
that they will be judged for making one choice in preference to another.  In
addition, supporting privacy in financial transactions can directly reduce some
forms of crime, such as domestic abuse~\cite{cashessentials2019}.

Irrevocable anonymity for payers is not incompatible with regulation, as might
be assumed.  Removing the ability to determine the payer in a transaction
channel does not entail removing the ability to determine the recipient, or
allowing transactions to take place outside the view of regulators more
generally.  In fact, privacy for payers complements regulations ensuring that
recipients are accountable for tax obligations, and a system that supports the
ability for payers to verify the identity of recipients can reduce fraud
without compromising their anonymity.  Finally, by requiring recipients of
digital currency to satisfy AML/KYC requirements, we can ensure that all
transacting parties, even payers, are within the security envelope of the set
of identified users, without linking specific users to the transactions in
which they served as payer.

\subsection{Property}

In many cultures, the right to private property is broadly considered to be a
foundational human right.  Money is no different.  English law, for example,
includes the idea of a \textit{chose in action} (a debt), which is something
whose value in a transaction indexes the legal right of a particular party to
transact it, and a \textit{chose in possession}, which is something whose value
in a transaction is intrinsic to holding it.

Cash might be a \textit{chose in action} in the sense that it represents an
obligation of the Bank of England, but it is also a \textit{chose in
possession} of its bearer as an instrument that can be held, owned, and
possessed.  Bank deposits are different; from the perspective of the beneficial
owner, they are claims on the assets of the bank, a \textit{chose in action}
only.  Bank deposits have also replaced cash, in a growing number of countries,
as the primary means by which individual persons make retail purchases, raising
the question: \textit{If users cannot own their own money, then what can they
own?}

Recently, legal scholars have considered the status of cryptoassets, and there
is a growing consensus among scholars of English law that simply treating them
as \textit{choses in action} is insufficient~\cite{norton2021,addleshaw2022}.
As it turns out, it is possible to have direct possession and control of
digital assets, and there exist technical mechanisms for exchanging digital
assets that do not require identity and accounting infrastructure.  More
saliently, the UK Parliament recently enacted the Electronic Trade Documents
Act, which explicitly offers a path to recognising the possession of certain
digital trade documents by their bearers, as well as the rivalrous transfer of
such documents from one party to another~\cite{uk2023}.  Such legal reasoning
could form the basis for recognising the rights of individual persons to
possess and control their own digital assets, including digital cash.

\subsection{Addressing shortcomings of existing retail payments
infrastructure}

Card payment platforms dominate the existing digital payments infrastructure
for retail payments.  As public infrastructure, these payment platforms have
important shortcomings.  For example, we know that the existence of merchant
fees means that merchants who charge the same price for cash payments as for
card payments implicitly allow customers who pay with cash to subsidise those
who pay with cards.  Although the merchant fees might apply only to the card
transactions, those fees amount to an economic tax on the transaction and
ultimately increase the equilibrium price for consumers.  Additionally, the
benefits from rewards programmes offered by card schemes accrue differently to
different classes of consumers.  As researchers at the Federal Reserve Bank of
Boston observed, ``[b]ecause credit card spending and rewards are positively
correlated with household income, the payment instrument transfer also induces
a regressive transfer from low-income to high-income households in
general''~\cite{schuh2010}.

In addition to intrinsic flaws with card payments, there are extrinsic problems
as well.  The largest card platforms, including Visa and Mastercard, are based
in the US, much to the chagrin of central banks and payment system regulators
in other countries.  Given the burgeoning systemic indispensability of card
payments to many economies around the world, it would be understandable for
businesses and governments of countries outside the US to seek strategic
autonomy, if for no other reason than to protect themselves from the
possibility that the US government might act adversely, for example, to
introduce tariffs, or to force card platforms to withdraw from certain markets
entirely.

For all of these reasons, it might be reasonable to ask whether card payments,
a twentieth century invention, remain fit for purpose in the twenty-first
century and beyond.  Central bank digital currency might offer a way to rethink
the role of card payments in the economy in general, as well as to provide a
public payment option for consumers in the digital economy in particular.

\section{The future of money}
\label{s:future}

Governments, central banks, and prominent institutions have taken a variety of
approaches to the future of money.  These approaches range from enthusiastic to
dismissive, from aggressive to passive, and from open to secretive.  At the
same time, not all stakeholders have a voice in this debate, and not all of the
universe of possible CBDC designs are being considered and represented with
equanimity and fairness.  Amidst the cacophony, some clear themes have emerged.
Collectively, these themes form the basis of the approaches taken by many of
the nations, their governments, and central banks whose currencies might one
day be minted in digital form, many of which reflect an echo chamber of
reverberating, unimpressive CBDC design properties that do not place the public
and their needs at the centre of the CBDC design debate as the ultimate
beneficiaries of this new public digital payment infrastructure.

Breaking through the noise requires building upon the research that exposes the
unsoundness in the foundations underpinning the prevailing proposals for the
design of CBDC.  For example, Hyoung-kyu Chey observed the difference between a
``state theory of money'', which posits that the foundational role of money is
its value as a unit of account, and a ``commodity theory of money'', which
posits that the foundational role of money is its value as a medium of
exchange~\cite{chey2023}.  While the prevailing CBDC proposals seem to assume
that the primary role of money is to serve as a unit of account, experience
with cryptocurrencies suggests that the need for CBDC arises instead from the
role of money as a medium of exchange, highlighting a mismatch between the
theoretical framework from which account-based CBDC proposals seem to arise and
the constraints facing retail consumers as they pay for goods and services in
the digital age.  Similarly, although some of the justifications given for the
proposed designs for CBDC reflect fears about the disintermediation of banks,
research by Bibi and Canelli suggests that such fears are
overwrought~\cite{bibi2024}.

The debate about the design of digital currency is not only a debate about what
form of payments are available to users. Rather, it is broader: The debate
about CBDC design is about framing and planning the financial infrastructure
that will power the digital economy, first domestically and in a retail
context, and then globally.  Now is the time for us to decide upon a vision for
a new ``Internet of value'', powered by payments, akin to the way early forms
of cash and banknotes that were created to address, among other things, the
need for individuals to buy and sell things without intermediation.  We should
encourage policymakers to take a step back from the press releases and ask:
What are the properties that digital money should have?  We suggest the
following desiderata for a public digital currency system:

\begin{itemize}

\item Is non-discriminatory by design.

\item Offers privacy by design for payers.

\item Allows consumers to directly possess money and control its ownership.

\item Supports an open architecture without the need for certified hardware.

\end{itemize}

A similar debate is taking place \textit{vis-\`a-vis} the digital economy in
general. The debates covered in this article and the motivations of central
banks, policy makers and financial services providers are diverse, and yet,
they do not centre around the needs and delivering the greatest value to users
and consumers. To the contrary, the consequences of the designs proposed by
teams in the US, UK, and EU are in fact detrimental to the interests,
preferences, protections, and rights of individuals and households that are
ultimately the consumers that would be expected to use a CBDC.  Specifically,
those individuals and households would not possess and control their own money,
but instead would merely have a contingent claim on assets possessed and
controlled by others.  Furthermore, those individuals and households would be
accountable to their custodians for how, where, when, and with whom they spend
their money.  Indeed, a vast chasm continues to grow between the set of
institutionally promoted CBDC design proposals and the set of alternative
models that must be publicly debated.

On one hand are privacy-thwarting and custodially-mediated designs that force
individuals into relationships, wherein every payment transaction that users
make is logged and linked to their identities by someone, somewhere, and
wherein users are forced to accept the involvement of intermediaries that can
control, censure, monitor, and individually restrict their use of money.  On
the other hand are private, non-custodial designs, for which it is technically
impossible to build profiles and restrictions for individual users or to force
users to ask permission from an authority when they spend their money.  When
framed in this way, as an evolutionary and iterative development for money and
payments, and as the plumbing for the digital economy, the need for
public debate and appreciation becomes apparent.

Part and parcel with this debate is the establishment of groundwork for the
desiderata that will govern the local and global retail payment infrastructure
in the first instance, and which will extend to other areas of finance and
value as they digitise, broaden, and become more integrated.  The CBDC
proposals seem to assume that all digital money must inexorably be held in
accounts, or that there is no technically viable alternative to having third
parties with institutional capabilities manage the money owned by natural
persons.  An argument on the basis of this assumption is casuistry, and it
would seem to imply that direct ownership of money, and by extension anything
of value that can be used as countertrade in an economic transaction, is a
thing of the past.  We reject such arguments.  For us, the best model for
digital currency would feature, in addition to our four desiderata, the
following fundamental characteristics:

\begin{itemize}

\item \textit{tokens} that can be possessed directly, rather than balances that
can only represent money that is possessed and controlled by others;

\item a \textit{security model} based upon data and compliance restrictions
associated with \textit{those who receive money} in a transaction, not those
who originate it, since every originator was, after all, once a receiver;

\item \textit{no holding limits,} not because individuals should be always be
allowed to receive more money, but because they should be able to hold money
on their own terms;

\item a \textit{scalable operating model} that allows transactions to be
processed in a distributed manner rather than via a small number of high-value
control points; and

\item support for \textit{open-source self-custody wallets} that offer
confidence to individuals and households that their possession and control of
money does not depend upon the performance or practices of third-party
\textit{de facto} custodians.

\end{itemize}

We suggest that the core desiderata and design characteristics of public
infrastructure should be agreed upon by consensus and should position the
individual user at the centre, taking their interests into account as
non-negotiable requirements and foundational constraints, rather than treating
them as easily curtailed pawns that the G7 and others have so casually set
aside in furtherance of a narrow set of interests.

\section*{Acknowledgements}

The authors acknowledge the support and encouragement of anonymous reviewers,
anonymous central bankers, the UCL Future of Money Initiative, and the Systemic
Risk Centre at the London School of Economics.

\end{document}